\documentclass[pra,twocolumn,aps,groupedaddress,showpacs]{revtex4}
\usepackage{amsmath,amsfonts,amssymb}
\usepackage{bbm,verbatim}

\def\ket#1{\left| #1\right>}

\newcommand{\Az}{A_z}
\newcommand{\Am}{A_-}
\newcommand{\Ap}{A_+}

\def\assign{\mathrel{\raise.095ex\hbox{:}\mkern-4.2mu=}}
\def\ngissa{\mathrel{=\mkern-4.2mu\raise.095ex\hbox{:}}}

\def\Hint{H_{\mathrm{int}}}
\def\Tint{T_{\mathrm{int}}}
\def\Hff{H_{\mathrm{ff}}}
\def\Hzz{H_{\mathrm{zz}}}

\def\up{\uparrow}
\def\dn{\downarrow}

\newcommand{\mr}[1]{\mathrm{#1}}
\newcommand{\oV}{\overline{V}}
\newcommand{\Eqref}[1]{Eq.~(\ref{#1})}
\renewcommand{\eqref}[1]{Eq.~(\ref{#1})}

\renewcommand{\url}[1]{\relax}
\def\urlprefix{\relax}

\usepackage{wrapfig}
\usepackage{graphicx}
\usepackage{graphics}

\def \beq {\begin{equation}}
\def \eeq {\end{equation}}
\def \ba {\begin{eqnarray}}
\def \ea {\end{eqnarray}}

\newcommand{\mean}[1]{\langle#1\rangle}

\def\ket#1{\left| #1\right>}

\def\avDAz{\overline{\Delta A_{z, \mr{est}}}^{(M)}}

\begin{document}

\title{Quantum measurement of a mesoscopic spin ensemble} \author{G. Giedke$^{1,4}$, J. M. Taylor$^2$, D. D'Alessandro$^3$,
  M. D. Lukin$^2$, and A. Imamo\u{g}lu$^1$} \affiliation{ $^1$
  Institut f\"ur Quantenelektronik, ETH Z\"urich,
  Wolfgang-Pauli-Stra{\ss}e
  16, 8093 Z\"urich, Switzerland \\
  $^2$ Department of Physics, Harvard University, Cambridge, MA 02138, USA\\
  $^3$ Department of Mathematics, Iowa State University, Ames, IA
  50011, USA\\
$^4$ Max-Planck--Institut f\"ur Quantenoptik, H.-Kopfermann-Str.,
85748 Garching, Germany
}

\begin{abstract}
  We describe a method for precise estimation of the polarization of a
  mesoscopic spin ensemble by using its coupling to a single two-level system.
  Our approach requires a minimal number of measurements on the two-level
  system for a given measurement precision. We consider the application of
  this method to the case of nuclear spin ensemble defined by a single
  electron-charged quantum dot: we show that decreasing the electron spin
  dephasing due to nuclei and increasing the fidelity of nuclear-spin-based
  quantum memory could be within the reach of present day experiments.
\end{abstract}
\maketitle

\section{Introduction}
\label{sec:introduction}

Decoherence of quantum systems induced by interactions with low-frequency
reservoirs is endemic in solid-state quantum information processing (QIP)
\cite{JPT+05,ICJ+05}. A frequently encountered scenario is the coupling of a
two-level system (qubit) to a mesoscopic bath of two-level systems such as
defects or background spins. The manifestly non-Markovian nature of
system-reservoir coupling in this scenario presents challenges for the
description of the long term dynamics as well as for fault tolerant quantum
error correction \cite{AHHH02,TeBu05}. The primary experimental signature of a
low-frequency reservoir is an unknown but slowly changing effective field that
can substantially reduce the ability to predict the system dynamics. A
possible strategy to mitigate this effect is to carry out a quantum
measurement which allows for an estimation of the unknown reservoir
field by controlled manipulation and measurement of the qubit. A precise
estimation of the field acting on the large Hilbert space of the reservoir
requires, however, many repetitions of the procedure: this constitutes a major
limitation since in almost all cases of interest projective measurements on
the qubit are slow \cite{EHW+04} and in turn will limit the accuracy of the
estimation that can be achieved before the reservoir field changes.

In this work, we propose a method for estimating an unknown quantum field
associated with a mesoscopic spin ensemble.  By using an incoherent version of
the \emph{quantum phase estimation algorithm}, \cite{Kit95,NC00} we show that
the number of qubit measurements scale linearly with the number of significant
digits of the estimation.  We only assume the availability of single qubit
operations such as preparation of a known qubit state, rotations in the
$xy$-plane, and measurement, of which only rotations need to be fast.  The
estimation procedure that we describe would suppress the dephasing of the
qubit induced by the reservoir; indeed, an interaction with the estimated field
leads to coherent unitary evolution that could be used for quantum control of
the qubit.  If the measurement of the reservoir observable is sufficiently
fast and strong, it may in turn suppress the free evolution of the reservoir
in a way that is reminiscent of a quantum Zeno effect.

After presenting a detailed description of the measurement procedure and
discussing its performance and limitations, we focus on a specific application
of the procedure for the case of a single quantum dot (QD) electron spin
interacting with the mesoscopic nuclear spin ensemble defined by the QD. It is
by now well known that the major source of decoherence for the electron-spin
qubits in QDs \cite{LoDi98} is the hyperfine interaction between the spins of
the lattice nuclei and the
electron~\cite{KLG02,MER02,SKL03,CoLo04,ErNa04,SSW04,DeHu03}.  A particular
feature of the hyperfine-related dephasing is the long correlation time
($t_c$) associated with nuclear spins.  This enables techniques such as
spin-echo to greatly suppress the dephasing~\cite{PJT+05}.  In \cite{CoLo04}
it was suggested to measure the nuclear field to reduce electron spin
decoherence times; precise knowledge of the instantaneous value of the field
would even allow for controlled unitary operations.  For example, knowledge of
the field in adjacent QDs yields an effective field gradient that
could be used in recently proposed quantum computing approaches with pairs of
electron spins~\cite{TED+05}.  Moreover, with sufficient control, the
collective spin of the nuclei in a QD may be used as a highly coherent
qubit-implementation in its own right~\cite{TML03,TIL03,TGC+04}.

\section{Phase estimation}

In the following we consider an indirect measurement scheme in which the
system under investigation is brought into interaction with a probe spin (a
two-level system in our case) in a suitably prepared state.  Measuring the
probe spin after a given interaction time $t$ yields information about the
state of the system. We assume the mesoscopic system evolves only slowly
compared to the procedure, and further that the measurement does not directly
perturb the system. In essence, we are performing a series of quantum
non-demolition (QND) measurements on the system with the probe spin.

We consider an interaction Hamiltonian of the form 
\begin{equation}
  \Hint = \hbar A_z\otimes S_z   \label{eq:Hint}
\end{equation}
which lends itself easily to a measurement of the observable $A_z$.
The QND requirement is satisfied for $[H_{\rm
  int},H_{\rm bath}] \rightarrow 0$.  The applicability of $\Hint$ in
situations of physical interest is discussed in
Sec.~\ref{sec:example}. 
 Given this interaction, the strategy to
measure $A_z$ is in close analogy to the so-called Ramsey
interferometry approach, which we now briefly review.

For example, an atomic transition has a fixed, scalar value for $A_z$
which corresponds to the transition frequency.  By measuring $A_z$ as
well as possible in a given time period, the measurement apparatus can
be locked to the fixed value, as happens in atomic clocks.  
The probe spin $S$ is
prepared in a state $\ket{+} = (\ket{\up} + \ket{\dn})/\sqrt{2}$.
It will undergo evolution under $H$ according to $U_t = \exp(-i t A_z
S_z)$.  After an interaction time $t$, the probe spin's state
will be 
\begin{equation}
  \cos(\Omega t) \ket{+} + i \sin(\Omega t) \ket{-} \label{e:class}
\end{equation}
where $\Omega = A_z / 2$ is the precession frequency for the probe
spin.  A measurement of the spin in the $\ket{\pm}$ basis yields a
probability $\cos^2(\Omega t)$ of being in the $\ket{+}$ state.
Accumulating the results of many such measurements allows one to
estimate the value for $\Omega$ (and therefore $A_z$).  In general,
the best estimate is limited by interaction time: for an expected
uncertainty in $A_z$ of $\Delta_0$ and an appropriate choice of $t$,
$M$ measurements with fixed interaction times $1/\Delta_0$ can
estimate $A_z$ to no better than $\sim \Delta_0 / \sqrt{M}$ (see
\cite{WBB+91} and references therein).

In our scenario, the situation is slightly different in that $A_z$ is
now a quantum variable.  For a state $\ket{s}$ in the Hilbert space of
the system $\mathcal{H}$ which is an eigenstate of $A_z$ with
eigenvalue $2 \Omega_s$, the coupling induces oscillations:
\begin{equation}
  U_t \ket{s} \ket{+}
= \ket{s}\left[\cos(\Omega_s t) \ket{+} + i \sin(\Omega_s t)
  \ket{-}\right] \ . \label{e:quant} 
\end{equation}
Thus, the probability to measure the probe spin in state $\ket{+}$
given that the system is in a state $\ket{s}$ is
$p(+|s)=\cos^2(\Omega_st)$ at time $t$, providing information about
which eigenvalue of $A_z$ is realized.  Comparing \Eqref{e:class} to
\Eqref{e:quant} indicates that the same techniques used in atomic
clocks (Ramsey interferometry) could be used in this scenario to
measure $\Omega_s$ and thus project the bath in some eigenstate of
$A_z$ with an eigenvalue of $\Omega_s$ to within the uncertainty of
the measurement.  

Beyond the Ramsey approach, there are several ways to extract this
information, which differ in the choice of interaction times $t_j$ and the
subsequent measurements.  The general results on quantum metrology of
\cite{GLM06} show, however, that the standard Ramsey scheme with fixed
interaction time $t$ is already optimal in that the scaling of the final
variance with the inverse of the total interaction time cannot be improved
without using entangled probe states. Nevertheless, the Ramsey scheme will not
be the most suitable in all circumstances. For example, 
we have assumed so far that
preparation and measurement of the probe spin is fast when compared to
$1/\Delta_0$.  However, in many situations with single quantum
systems, this assumption is no longer true, and it then becomes desirable
to minimize the number of preparation/measurement steps in the
scheme. 

\section{The measurement scheme}
\label{sec:scheme}
\begin{figure}
\includegraphics[width=3.0in]{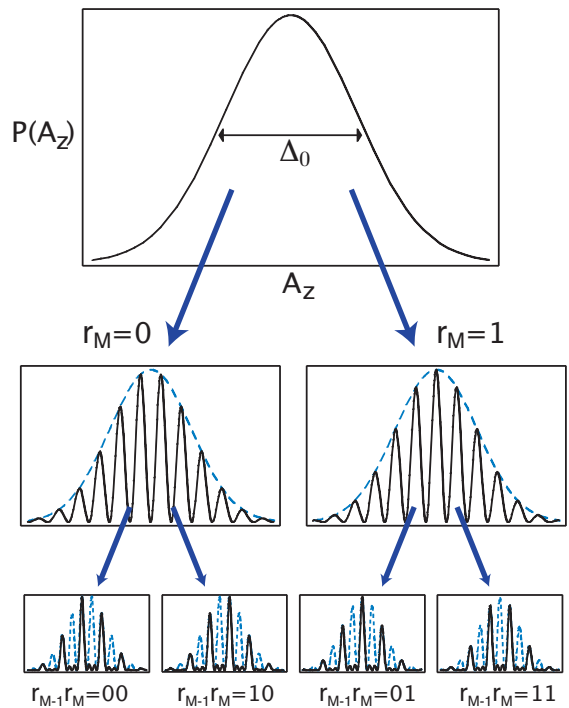}
\caption{(Color online)
  Illustration of the first two steps of the measurement procedure.
  The original distribution $P(A_z)$, with rms width $\Delta_0$ is shown at the top.  
  After the
  first measurement, with result $r_M = 0,1$, the conditional
  distribution (middle plots) reflects the knowledge of the least significant bit.
  The next measurement result $r_{M-1}$ further reduces the distribution (bottom plots).
\label{f1}}
\end{figure}
We now show that by varying the interaction time and the final
measurements such that each step yields the maximum information about
$\Omega_s$, we can obtain the same accuracy as standard Ramsey
techniques with a similar interaction time, but only a {\em logarithmic}
number of probe spin preparations and measurements.  As a trivial
case, if $A_z$ had eigenvalues $0$ and $1$ only, then measuring the
probe in the $\pm$-basis after an interaction time $t_1=\pi$, we find
$+(-)$ with certainty, if the system are in an $A_z=0(1)$-eigenstate;
if they were initially in a superposition, measuring the probe
projects the system to the corresponding eigenspaces.  We can extend
this simple example (in the spirit of the quantum phase estimation
algorithm \cite{Kit95,NC00} and its application to the measurement of
a classical field \cite{VaMi04}) to implement an $A_z$-measurement by
successively determining the binary digits of the eigenvalue.  We
start with the ideal case, then generalize to a more realistic scenario.

\subsection{Ideal case}
If all the eigenvalues of $A_z$ are an integer multiple of some known number
$\alpha$ and bounded by $2^M\alpha$, then this procedure yields a perfect
$A_z$-measurement in $M$ steps: let us write all eigenvalues as
$2\Omega_s=\alpha2^M\sum_{l=1}^Ms_l2^{-l}$. The sum we denote by $s$ and also
use the notation $s = 0.s_1s_2\dots s_M$. 
Starting now with an interaction time $t_1=\frac{\pi}{\alpha}$, we have
$\Omega_st_1=s_M\frac{\pi}{2}\,\mr{mod}\,\pi$. Hence the state of the probe
electron is flipped if and only if $s_M=1$. Therefore measuring the probe
electron in state $+(-)$ projects the nuclei to the subspace of even(odd)
multiples of $\alpha$ (see Fig.~\ref{f1}).  We denote the result of the first measurement by
$r_M=0(1)$ if the outcome was ``$+(-)$''. All the higher digits have no effect
on the measurement result since they induce rotations by an integer multiple
of $\pi$ which have no effect on the probabilities $p(\pm|s)$. 

To measure the higher digits, we reduce the interaction time by half
in each subsequent step: $t_{j+1} = 2^{-j}t_1$ until we reach $t_M =
\frac{\pi}{\alpha}$ in the final and shortest step.  For $j>1$ the
rotation angle $\Omega_st_j$ (mod $\pi$) in the $j$th step does not
only depend on the $j$th binary digit of $s$ but also on the previous
digits (which have already been measured, giving results $r_{M+1-l} =
s_{M+1-l}, l=1,\dots,j-1$). The angle $\Omega_st_j$ (mod $\pi$) is
given by $s_{M+1-j}\frac{\pi}{2} + \varphi_j$ with $\varphi_j =
\frac{\pi}{2}\sum_{l=1}^{j-1}r_{M+1-l}2^{l-j}$, where we have used the
results $r_l$ already obtained. This over-rotation by the angle
$\varphi_j$ can be taken into account in the choice of the measurement
basis for the $j$th step: if the $j$th measurement is performed in a
\emph{rotated basis} $\ket{\pm_j}$ that is determined by the previous
results $r_{l}$, namely
\begin{subequations}
  \label{eq:3}
  \begin{eqnarray}
      \ket{+_j} &\assign& \cos\varphi_j\ket{+}-i\sin\varphi_j\ket{-},\\
      \ket{-_j} &\assign& \sin\varphi_j\ket{+}+i\cos\varphi_j\ket{-},
  \end{eqnarray}
\end{subequations}
then the $j$th measurement yields ``$+$'' ($r_{M+1-j}=0$) if
$s_{M+1-j}=0$ and ``$-$'' ($r_{M+1-j}=1$) otherwise. Thus, after $M$
measurements we obtain  $r_l=s_l, \forall l=1,\dots,M$ and have
performed a complete measurement of $A_z$ (where the number
$M$ of probe particles used is the smallest integer such that $2^M\geq
A_z/\alpha$). 

Before proceeding, we note that the proposed scheme is nothing but an
``incoherent'' implementation of the quantum phase estimation
algorithm: As originally proposed, this algorithm allows measurement
of the eigenvalue of a unitary $U$ by preparing $M$ qubits (the
control-register) in the state $\ket{+}^{\otimes M}$ (i.e., the
equal superposition of all computational basis states $\ket{j},
j=0,\dots,2^M-1$) and performing controlled-$U^{2^{j-1}}$ gates
between the $j$th qubit and an additional register prepared in an
eigenstate $\ket{s}$ of $U$ with $U\ket{s} = e^{i2\pi s}\ket{s}$. The
controlled-$U$ gates let each computational basis state
acquire a $s$-dependent phase: $\ket{l}\mapsto e^{i2\pi ls}\ket{l}$.
Then the inverse quantum Fourier transformation (QFT) is performed on
the control register, which is then measured in the computational basis,
yielding the binary digits of $s$. Performing the QFT is still a
forbidding task, but not necessary here: 
the sequence of measurements in the rotated basis $\ket{\pm_j}$
described above is in fact an implementation of the combination of QFT
and measurement into one step. This was previously suggested in
different contexts \cite{GrNi96,PaPl00,ToNa04}.

\subsection{Realistic case}
In general, there is no known $\alpha$ such that all eigenvalues $s$
of $A_z$ are integer multiples of $\alpha$.  Nevertheless, as
discussed below, the above procedure can still produce a very accurate
measurement of $A_z$ if sufficiently many digits are measured.
Now we evaluate the performance of the proposed measurement scheme in
the realistic case of non-integer eigenvalues. Since here we are
interested in the fundamental limits of the scheme, we will for now
assume all operations on the probe qubit (state preparation,
measurement, and timing) to be exact; the effect of these imperfection
is considered in Sec.~\ref{sec:errors}. Without loss of
generality, let $A$ and $0$ denote the largest and smallest
eigenvalues of $A_z$, respectively
\footnote{
In practice one may want to make use of prior knowledge about the
state of the system to reduce the interval of possible eigenvalues
that need to be sampled. Hence $A$ may be understood as an effective
maximal eigenvalue given, e.g., by the expectation value of $A_z$ plus
$f$ standard deviations. The values outside this range will not be
measured correctly by the schemes discussed, but we assume $f$ to be
chosen sufficiently large for this effect to be smaller than other
uncertainties.} 
and choose
$\alpha=2A$ such that the eigenvalues of $A_z/\alpha$ are all
$\in[0,1/2]$. These are the eigenvalues $s$ we measure in the
following.

The function from which all relevant properties of our strategy can be
calculated is the conditional probability $p_M(R|s)$ to obtain (after
measuring $M$ electrons) a result $R=0.r_1r_2\dots r_M$ given that the system
was prepared in an eigenstate with eigenvalue $s$. The probability to measure
$R$ is given by the product of the probabilities to measure $r_{M+1-j}$ in the
$j$th step, which is $\cos^2(\Omega_st_{j}-\varphi_j+r_{M+1-j}\frac{\pi}{2}) =
\cos^2(\pi [s-R]2^{M-j})$. Hence
\begin{equation}
  \label{eq:5}
  p_M(R|s) = \prod_{k=0}^{M-1} \cos^2(\pi[s-R]2^k),
\end{equation}
see also \cite{VaMi04}. This formula can be simplified by repeatedly using
$2\sin x = \sin (x/2)\cos(x/2)$ to give
\begin{equation}
  \label{eq:6}
p_M(R|s) =  \left(\frac{\sin(2^M\pi[s-R])}{2^M\sin(\pi[s-R])}\right)^2.  
\end{equation}
Assume the nuclei are initially prepared in a state $\rho$ with prior
probability $p(s)$ to find them in the eigenspace belonging to the eigenvalue
$s$. After the measurement, we can update this distribution given our
measurement result. We obtain, according to Bayes' formula:  
\begin{equation}
  \label{eq:7}
  p_M(s|R) = \frac{p_M(R|s) p(s)}{\sum_s
    p_M(R|s)p(s)},
\end{equation}
with expectation value denoted by $\bar{s}_R$. 

\section{Performance of the scheme}

As the figure of merit for the performance of the measurement scheme
we take the improvement of the average uncertainty in $A_z$  of the updated distribution 
\begin{equation}
  \label{eq:4}
\avDAz \assign \sum_R p_M(R)\sqrt{\sum_s (s-\bar
  s_R)^2    p_M(s|R)}
\end{equation}
over the initial uncertainty $\Delta_0 = \Delta A_z^{(0)}$. 
An upper bound to $\avDAz$ is given by the square root of the
average variance $\oV$ 
\begin{equation}
  \label{eq:8}
   \oV_M \assign \sum_R p_M(R)\sum_s (s-\bar s_R)^2    p_M(s|R),
\end{equation}
as easily checked by the Cauchy-Schwarz inequality. 
We now show that $\oV_M \leq 2^{-M}$.  We replace $\bar s_R\to \tilde
R= \mr{min}\{R,1-R\}$; we can use any such replacement to obtain an
upper bound, as the expectation value $\bar x=\sum_xp(x)x$ minimizes
$v(y)=\sum_x p(x)(x-y)^2$. This choice means that measurement results
$R>1/2$ are interpreted as $1-R$, which is appropriate since the
scheme does not distinguish the numbers $s=\delta$ and $s'=1-\delta$
and due to the choice of $\alpha$ only $s\in[0,1/2]$ occur. Thus 
\begin{eqnarray*}
  \oV_M &\leq& \sum_s p(s)\sum_R (s-\tilde R)^2
  p_M(R|s)\ngissa \sum_s p(s) \oV_M(s).
\end{eqnarray*}
The terms $\oV_M(s)$ can be shown\footnote{For this we make use of
  \eqref{eq:6} and $(\sin x)/x \geq 1-\pi^{-1}|x|$ (for $|x|\leq\pi$)
  to bound all terms of the sum over $R$.} to be $\leq b_V\, 2^{-M}$
with $b_V = (1+2^{-M})/2$. This means that performing $M$ measurements
yields a state with $A_z$-uncertainty $\Delta A_{z,\mr{est}}\leq
\alpha 2^{-M/2}$. For example, we need about $13$ interactions with
the probe spin to reach the $1\%$-level in $\Delta A_{z,
  \mr{est}}/\alpha$
and about $7$ more for 
every additional factor of $10$. 

The overall procedure requires a total time $T_M=\sum_{j=1}^M (t_j+\tau_m) =
2t_1(1-2^{-M})+M\tau_m$, which is an interaction time (determined mainly by the
time $t_1 = \frac{\pi}{\alpha}2^M=\pi(2f\Delta_0)^{-1}2^M$ needed 
for the least significant digit probed) and the time to make $M$
measurements ($\tau_m$ is the time to make a single measurement).  We
obtain for the average uncertainty an upper bound in terms of the
interaction time $T_{\rm int} = T_M - M \tau_m$ needed:
\begin{equation} 
  \label{eq:9}
  \frac{\Delta A_{z, \mr{est}}}{\Delta_0} \leq \sqrt{\frac{\pi f}{\Delta_0 \Tint}}.
\end{equation}
Immediately the similarity with standard atomic clock approaches is
apparent, as the uncertainty decreases with the square root of the
interaction time.  However, while for an atomic clock scheme, in which
the interaction time per measurement is kept fixed to $\sim \pi/(f
\Delta_0)$, the total time to reach the precision of \Eqref{eq:9} is $T^{\rm
  r}_M = 2 t_1 + \tau_m 2^M$.  For our method the measurement time
is reduced dramatically by a time $T^{\rm r}_M - T_M = \tau_m (2^M -
M)$. In this manner our approach requires a polynomial, rather than
exponential, number of measurements for a given accuracy, though the
overall interaction time is the same for both techniques.  

It may be remarked that even the scaling in interaction time differs
significantly if other figures of merit are considered. For example,
our scheme provides a \emph{square-root speed-up} in $\Tint$ over the
standard Ramsey scheme if the aim is to maximize the information gain
or to minimize the confidence interval \cite{masanes02notes}.

\section{Errors and Fluctuations in $A_z$}
\label{sec:errors}

Up until now we have considered an idealized situation in which the
value of $A_z$ does not change over the course of the measurement and
in which preparation and measurement of the probe system work with
unit fidelity. Let us now investigate the robustness of our scheme in
the presence of these errors.

\subsection{Preparation and Measurement Errors} 
\begin{figure}
\includegraphics[width=3.3in]{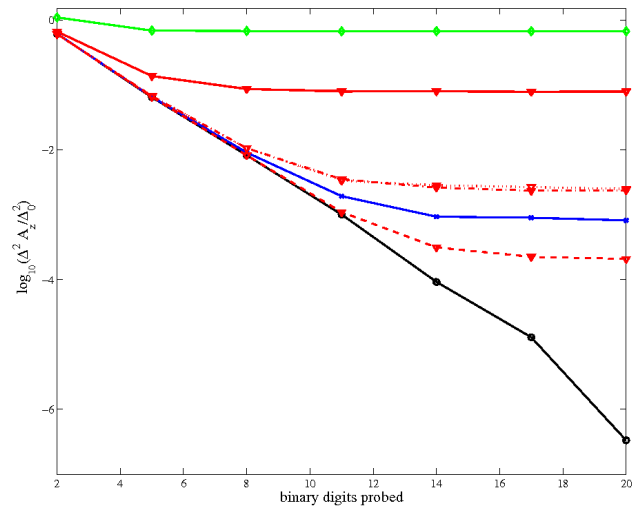}
\caption{(Color online)
Estimation in the presence of preparation and measurement error $p$: the
logarithm of the improvement $\Delta A_{z,{\rm est}}/\Delta_0$ is plotted
versus the number $M$ of binary digits measured. The solid lines represent
different error rates ($p=0$: black circles; $p=10^{-4}$ blue crosses;
$p=10^{-2}$: red triangles; and $p=10^{-1}$: green diamonds).\\
The broken red curves (triangles) demonstrate the benefit of simple error
correction (for 
$p=10^{-2}$): strategy I (3 repetitions per digit, use majority result; 
either for all digits or only for leading $M/2$ digits): (dash-dotted, dotted
-- almost undistinguishable); strategy II (increase number of repetitions
for more significant digits to 7 repetitions for leading $M/8$ digits, 5 for
next $M/8$, and 3 for next $M/4$): dashed.  We see that the latter provides
a better accuracy than the uncorrected $p=10^{-4}$ curve.
\label{f2}}
\end{figure}

By relying upon a small number of measurements, the scheme we described
becomes more susceptible to preparation and measurement errors.  An error in
the determination of the $k$th digit leads to an increase of the error
probability in the subsequent digits.  This error amplification leads to
a scaling of the final error of $\sqrt{p}$, where $p$ is the probability of
incurring a preparation or measurement error in a single step. We confirm this
with Monte Carlo simulations of the measurement procedure (Fig.~\ref{f2}a),
leading to an asymptotic bound:
\begin{equation}
  \label{eq:13}
  \Delta A_{z, \mr{est}} \leq (2 f \Delta_0)\sqrt{p}.
\end{equation}
Standard error correction (EC) techniques can be used to overcome this problem.
E.g., by performing three measurements for each digit and using majority vote,
the effective probability of error can be reduced to $\sim 3p^2$ -- at the
expense of tripling the interaction time and number of measurements.  While this
may look like a big overhead, it should be noted that the
scheme can be significantly improved: the least significant digits do not
require any EC.  For them, the scheme gives noisy results
even for error rate $p=0$ due to the undetermined digits of $A_z$, this
does not affect the most significant $M/2$ digits. This indicates that it may
be enough to apply EC for the leading digits.  

As can be seen from Fig.~\ref{f2}b, this simple EC strategy provides a
significant improvement in the asymptotic $\Delta A_z$.  This is
hardly changed, when EC is applied only to the leading half of the
digits. Thus only twice as many measurements (and an additional
interaction time $\sim 2^{M/2+2}$ which is $\ll T_M$) is needed for an
order-of-magnitude improvement in $\Delta A_z$.  By repeating the
measurement of more important digits even more often, the effect of
technical errors can be reduced even further, as confirmed by Monte
Carlo simulations (Fig.~\ref{f2}b). We also note that further
improvements (beyond $M$ digits) can be achieved by this technique.
In essence, choosing a digital approach to error correction for our
digital technique yields substantially better performance than
adapting the digital technique to an analog approach.

\subsection{Estimation of bath decorrelation errors}
\label{sec:fluctuations}
In practice, internal bath dynamics will lead to fluctuations in $A_z$, such
that $A_z(t) \neq A_z(t')$ for times $t$ and $t'$ that are
sufficiently different.  Furthermore, apparatus errors, as outlined
above, lead to errors in our measurement procedure.  We will assume
that the variations of $A_z$ are slow over short time intervals,
allowing us to approximate the $M$ bit measurement process as a
continuous measurement over the time $T_M$ with some additional noise
with variance $\approx \Delta A_{z,{\rm est}} (M)^2 \ll \Delta_0^2$.  Then we will
find the expected difference in our measurement result and the value
of $A_z$ at a later time.

Under the above approximations the value of the $k$th such measurement (where
a complete set of $M$ bits takes a time $T_M$ and the $k$th such measurement
ends at time $t_k$) is 
\beq
m_k = \frac{1}{T_M} \int_{t_k-T_M}^{t_k} A_z(t) dt + G_k
\eeq
where the noise from measurement is incorporated in the stochastic noise variable $G_k$ with $\mean{G_k G_{k'}} \approx \delta_{kk'} \Delta A_{z,{\rm est}} (M)^2$.  We can estimate $A_z$ at a later time, and find the variance of this estimate from the actual value:
\begin{widetext}
\ba
\bar{V}_M(t>t_k) &=& \mean{[m_k-A_z(t)]^2} \nonumber \\
 & = & \mean{G_k^2} + \Delta_0^2 
+  \frac{1}{2 T_M^2} \int_{t_k-T_M}^{t_k}  \int_{t_k-T_M}^{t_k} \mean{\{A_z(t'), A_z(t'')\}_+} dt'' dt'   
-\frac{1}{T_M}  \int_{t_k-T_M}^{t_k} \mean{\{A_z(t') ,A_z(t)\}_+} dt' \nonumber
\ea

If we assume $A_z$ is a Gaussian variable with zero mean, described by a spectral function $S(\omega)$ (i.e., 
$\mean{A_z(t) A_z(t+\tau)} = \int_{-\infty}^{\infty} S(\omega) e^{i \omega \tau} d\omega$), then
\beq
\bar{V}_M(t) =  \Delta A_{z,{\rm est}} (M)^2 + \Delta_0^2  
+\frac{1}{T_M^2} \int_{-\infty}^{\infty} S(\omega) \frac{\sin^2(T_M \omega/2)}{(\omega/2)^2} d\omega  
-\int_{-\infty}^{\infty} S(\omega)
\frac{\sin[(t-t_k+T_M)\omega]-\sin[(t-t_k)\omega]}{T_M \omega/2}
d\omega \nonumber 
\eeq
\end{widetext}
For $A_z$ that fluctuates slowly in time and corresponds to a non-Markovian, low frequency noise, the second moment of $S(\omega)$ converges.  We define:
\beq
\frac{1}{t_c^2} = \frac{1}{\mean{A_z^2}} \int_{-\infty}^{\infty} S(\omega) \omega^2 d\omega\ .
\eeq
When 
$T_M, t-t_k+T_M \ll t_c$, 
we may expand the sine terms in the integrals.  Taking $t = t_k + T_M$, the
expected variance to order 
$(T_M/t_c)^2$ 
is
\beq
\bar{V}_M(t) \approx \Delta A_{z,{\rm est}} (M)^2 +  \Delta_0^2
\left[\frac{T_M}{t_c}\right]^2\left(\frac{7 }{3} - \frac{1}{12}\right) \label{e:decor}
\eeq 
As an example case, we consider as realistic parameters 
$t_c\approx 1$ms 
and
$T_M = 16\ \mu$s with $\Delta A_{z,{\rm est}}^2/\Delta_0^2 = 0.025^2$.  These parameter choices are described in detail in Sec.~\ref{sec:example}.
We find that our variance $16\mu$s {\em after} the measurement is
approximately $0.035^2 \Delta_0^2$ with equal contributions from the
measurement noise and from the bath decorrelation.  Substantially
faster decorrelation would dominate the noise in the estimate, and
render our technique unusable.

In the limit of slow decorrelation, this approach would allow one to
use the (random) field $A_z$ to perform a controlled unitary of the
form $\exp(-i m_k S_z \tau)$ at a time $t$, with a fidelity
\begin{equation} 
F=\exp( - \mean{(\int_{t-\tau}^t A_z(t') dt'-m_k)^2}/4) 
\end{equation}
For example, a $\pi$ rotation around the probe spins' $z$ axis would
have a fidelity $\approx 1 - \bar{V}_M(t) \pi^2/\Delta_0^2$, or 0.998
for the above parameters.

We remark that this approach for estimation in the presence of bath
fluctuations is not optimal (Kalman filtering~\cite{Kal60} would be more
appropriate for making an estimation of $A_z$ using the measurement results).
Furthermore, it does not account for the non-linear aspects of our measurement
procedure, nor does it incorporate any effect of the measurement on the {\em
  evolution} of the bath (e.g., quantum Zeno effect).  More detailed
investigations of these aspects of the process should be considered in an
optimal control setting.  Nonetheless, our simple analysis above indicates
that slow decorrelation of the bath will lead to modest additional error in
the estimate of $A_z$.

\section{Example: Estimating collective nuclear spin in a quantum dot}
\label{sec:example}

Now we apply these general results to the problem of estimating the
collective spin of the lattice nuclei in a QD. 

The interaction
of a single electron spin in a QD with the spins of the lattice nuclei
$\vec{I}_j$ is described by the Fermi contact term \cite{SKL03}
\begin{equation}
  \label{eq:12}
  \vec{S}\cdot\sum_j\alpha_j\vec{I}_j,
\end{equation}
where the sum in \Eqref{eq:12} runs over
all the $N$ lattice nuclei. The $\alpha_j$ are constants
describing the coupling of the $j$th nuclear spin with the electron.
They are proportional to the modulus squared of the electron
wave function at the location of the $j$th nucleus and are normalized
such that $\sum_j\alpha_j I^{(j)}=A$, which denotes the hyperfine
coupling strength. 

Due to the small size of the nuclear Zeeman energies, the nuclei are
typically in a highly mixed state even at dilution refrigerator
temperatures. This implies that the electron experiences an effective
magnetic field (Overhauser field, $\vec{B}_{\rm nuc}$) with large
variance, reducing the fidelity of quantum memory and quantum
gates. This reduction arises both from the inhomogeneous nature of the
field ($\vec{B}_\mr{nuc}$ varies from dot to dot) \cite{LAO+05} and the
variation of $\vec{B}_\mr{nuc}$ over time due to nuclear-spin dynamics
(even a single electron experiences different field strengths
over time, implying loss of fidelity due to time-ensemble averaging).

In a large external magnetic field in the $z$-direction the spin flips
described by the $x$ and $y$ terms are suppressed and -- in the interaction
picture and the rotating wave approximation -- the relevant Hamiltonian is of
the type given in \Eqref{eq:Hint}, where $A_z$ is now the collective nuclear
spin operator
\begin{equation}
  \label{eq:1}
  \Az = \sum_{j=1}^N \alpha_jI_z^{(j)},
\end{equation}
which gives the projection of the Overhauser field along the external field
axis by $B_{\mr{nuc},z} = \hbar A_z/g^*$. Before continuing, let us remark
here, that one can expect to obtain an effective coupling of the type
\Eqref{eq:Hint} in a similar fashion as a good approximation to a general
spin-environment coupling $\vec{S}\cdot\vec{A}$, whenever the computational
basis states of the qubit are non-degenerate (as guaranteed in the system
studied here by the external field) and the coupling to the environment is
sufficiently weak such that bit-flip errors are detuned.

To realize the single-spin operations needed for our protocol -- preparation,
rotation, and read-out -- many approaches have been suggested as part of a
quantum computing implementation with electron spin qubits in QDs using either
electrical or optical control (see, e.g., \cite{CCGL05} for a recent review).

The experimental progress towards coherent single spin manipulation has been
remarkable in recent years.  In particular, the kind of operations needed for
our protocol have already been implemented in different settings: For
self-assembled dots, state preparation with $F\geq 0.99$ has been realized
\cite{ADB+06}, while for electrically defined dots, single-spin measurement
with a fidelity of $F\geq0.72$ was reported \cite{HWV+05}. In the double-dot
setting \cite{PJT+05}, all three operations have recently been demonstrated,
and we estimate the combined fidelity to be $F \geq 0.7$.

As can be seen from Fig.~\ref{f2}, at the level of $1\%$ accuracy of state
preparation, rotation and read-out, the proposed nuclear spin measurement
should be realizable. As discussed in many specific proposals \cite{CCGL05}
these error rates appear attainable in both the transport and the optical
setting. Apart from single qubit operations, our proposal also requires
precise control of the interaction time.  Fast arbitrary wave form generators
used in the double-dot experiments, have time resolutions better than 30
ps\footnote{J. R.  Petta, Private communication} and minimum step sizes of 200
ps, which translates into errors of a few percent in estimating $A_z$ with
initial uncertainties of order 1 ns$^{-1}$.  Uncertainties of this
order are expected for large QDs ($N\sim 10^6$) even if they are
unpolarized and for smaller ones at correspondingly higher
polarization (see below).

For GaAs and InAs QDs in the single electron regime,
$A\sim50-200\mr{ns}^{-1}$ and $N\sim10^4-10^6$.  The uncertainty
$\Delta_0^2 = \langle \Az^2 \rangle - \langle \Az \rangle^2 =
(T_2^*)^{-2}$ determines the inhomogeneous dephasing time $T_2^*$
\cite{CoLo04}. Especially at low polarization $P$, this uncertainty is
large $\Delta_0^2\approx A^2(1-P^2)/N$, and without correction $\Hzz$
leads to fast inhomogeneous dephasing of electron spin qubits: 
$T_2^*\gtrsim 10$ ns has been observed~\cite{BSG+05,JPT+05,KFE+05}.
However, as $\Az$ is slowly varying~\cite{MER02,KLG02,PJT+05}, it may
be estimated, thereby reducing the uncertainty in its value and the
corresponding dephasing. This is expected to be particularly effective, when
combining estimation with recent progress in polarizing the nuclear spin
ensemble \cite{BSG+05,EKL+05,LMBI06}.  

In a QD system such as \cite{EHW+04}, with $\tau_m\simeq1\mu$s and
for $1/\Delta_0\simeq10$ns we can estimate 8 digits ($M = 8$) (improving
$\Delta A_{z, \mr{est}}$ by a factor of at least $16$) in a total time
$T_M=16\mu$s. In contrast, a standard atomic clock measurement scheme would
require a time $\simeq280\mu$s.

We now consider limits to the estimation process, focusing on expected
variations of $\Az$ due to nuclear spin exchange and preparation and
measurement errors.  Nuclear spin exchange, in which two nuclei switch
spin states, may occur directly by dipole-dipole interactions or
indirectly via virtual electron spin flips.  Such  flips lead
to variations of $\Az$ as spins $i$ and $j$ may have $\alpha_i \neq
\alpha_j$. 

The dipole-dipole process, with a $1/r^3$ scaling, may be approximated by a
diffusive process at length scales substantially longer than the lattice
spacing~\cite{Sli80,DeHu03}.  The length scale for a spin at site $i$ to a
site $j$ such that $\alpha_j \approx \alpha_i$ is not satisfied is on the order
of the QD radius (5-50 nm); for diffusion constants appropriate for
GaAs~\cite{Pag82}, the time scale for a change of $\Az$ comparable to $\Delta
\Az$ by this process is $\sim 0.01-10$s.

However, nuclear spin exchange mediated by virtual electron spin flips may be
faster.  This process is the first correction to the rotating wave
approximation, and is due to the (heretofore neglected) terms in the contact
interaction, $\Hff = \frac{\hbar}{2}(A_+S_-+A_-S_+)$, which are suppressed to
first order by the electron Larmor precession frequency $\epsilon_z$.  These
have been considered in detail
elsewhere~\cite{MER02,CoLo04,SSW05,YLS05,DeHu06,TPJ+06}.  Using perturbation
theory to fourth order, the estimated decorrelation time for $\Az$ is
$t_c^{-1} = A^2/(\epsilon_z N^{3/2})$, giving values $0.1-100$ms$^{-1}$ for
our parameter range \cite{TPJ+06}.  Taking $t_c =1$ ms, we may estimate the
optimal number of digits to measure. Using \Eqref{e:decor}, the best
measurement time is given by $T_M \sim t_c / 2^{M/2}$ and  for the values 
used
above, $M \approx 10-11$ is optimal.  We note as a direct corollary that our
measurement scheme provides a sensitive probe of the nuclear spin dynamics
on nanometer length scales.

We now consider implications of these results for improving the
performance of nuclear spin ensembles, both as quantum
memory~\cite{TIL03} and as a qubit~\cite{TGC+04}.  The dominant error
mechanism is the same as for other spin-qubit schemes in QDs:
uncertainty in $\Az$.  The proposed measurement scheme alleviates this
problem. However, the nuclear spin ensembles operate in a subspace of
collective states $\ket{0}$ and $\ket{1}$, where the first is a ``dark
state'', characterized by $\Am \ket{0} = 0$ (and the second is
$\propto \Ap \ket{0}$, where $A_\pm = \sum_j\alpha_j I_\pm^{(j)}$).
Thus $\ket{0}$ is an $\Am$ eigenstate and cannot be an $\Az$
eigenstate when $\alpha_k \neq {\rm const}$ (except for full
polarization).  Therefore, the measurement [which essentially projects
to certain $A_z$-''eigenspaces'' ($A_z\in[a-\Delta a,a+\Delta a]$)]
moves the system out of the computational space, leading to leakage
errors.  The incommensurate requirements of measuring $\Az$ and using
an $\Am$ eigenstate place a additional restriction on the precision of
the measurement. The optimal number of digits can be estimated in
perturbation theory, using an interaction time $\Tint\approx 2t_1$ and
numerical results \cite{TIL03} on the polarization dependence of
$\Delta_0$. We find that for high polarization $P>90\%$ a relative error of
$\Delta a/a \lesssim1\%$ is achievable.

\section{Conclusions}
\label{sec:conclusions}

We have shown that a measurement approach based on quantum phase estimation
can accurately measure a slowly varying mesoscopic environment coupled to a
qubit via a pure dephasing Hamiltonian.  
By letting a qubit interact for a sequence of well controlled times and
measuring its state after the interaction, the value of the dephasing variable
can be determined, thus reducing significantly the dephasing rate.

The procedure requires fast single qubit rotations, but can tolerate
realistically slow qubit measurements, since the phase estimation approach
minimizes the number of measurements. Limitations due to measurement and
preparation errors may be overcome by combining our approach with standard
error correction techniques. Fluctuations in the environment can also be
tolerated, and our measurement still provides the basis for a good estimate,
if the decorrelation time of the environment is not too short.

In view of the implementation of our scheme, we have considered the hyperfine
coupling of an electron spin in a quantum dot to the nuclear spin ensemble Our
calculations show that the Overhauser field in a quantum dot can be accurately
measured in times shorter that the nuclear decorrelation time by shuttling
suitably prepared electrons through the dot.  Given recent advances in
electron measurement and control \cite{EHW+04,PJT+05} this protocol could be
used to alleviate the effect of hyperfine decoherence of electron spin qubits
and allow for detailed study of the nuclear spin dynamics in quantum dots.
Our approach complements other approaches to measuring the Overhauser field in
a quantum dot that have recently been explored \cite{KCL06,SBGI06}. 

While we discussed a single electron in a single quantum dot, the method can
also be applied, with modification to preparation and measurement procedures
\footnote{In the strong field case a two-level approximation for the spin
  system is appropriate \cite{CoLo05}.  Preparing superpositions in the
  $S_z=0$ subspace such as the singlet, and measuring in this basis as
  well~\cite{PJT+05}, the scheme would measure the $z$-component of the
  nuclear spin difference between the dots. To measure the total Overhauser
  field, superpositions of the $S_z\not=0$ triplet states have to be used.},
to the case of two electrons in a double dot \cite{KFE+05,JPT+05,CoLo05}.  

As we have seen, the Hamiltonian \Eqref{eq:Hint} can serve as a good
approximation to more general qubit-environment coupling in the case of weak
coupling and a non-degenerate qubit. Therefore, we expect that this technique
may find application in other systems with long measurement times and slowly
varying mesoscopic environments.

\begin{acknowledgments}
  J.M.T. would like thank the quantum photonics group at ETH for their
  hospitality.  The authors thank Ignacio Cirac and Guifr\'e Vidal for sharing
  their notes on the performance of the QFT scheme for different figures of
  merit.  The work at ETH was supported by NCCR Nanoscience, at Harvard by
  ARO, NSF, Alfred P. Sloan Foundation, and David and Lucile Packard
  Foundation, and at Ames by the NSF Career Grant ECS-0237925, and at MPQ by
  SFB 631.
\end{acknowledgments}


\end{document}